\documentclass[aps,prc,twocolumn,showpacs,amsmath,preprintnumbers,amssymb,groupedaddress,superscriptaddress]{revtex4}
\usepackage{bbm}

\usepackage{graphicx}
\usepackage{dcolumn}
\usepackage{bm}
\usepackage{multirow}
\usepackage{hhline}

\begin{document}

\preprint{}

\title{Signature of QCD critical point: Anomalous transverse velocity dependence of antiproton-proton ratio}

\author{Xiao-Feng Luo}
\thanks{Corresponding author: xfluo@lbl.gov}

\author{Ming Shao}
\author{Cheng Li}
\author{Hong-Fang Chen}
\affiliation{University of Science and Technology of China, Hefei,
Anhui 230026, China}
\date{\today}

\begin{abstract}
We formulate the QCD critical point focusing effect on transverse
velocity ($\beta_{t}$) dependence of antiproton to proton
($\bar{p}/p$) ratio, which was recently proposed by Asakawa {\it et
al.} as an experimental signature of QCD critical point in high
energy heavy ion collisions (HICs). For quantitative analysis,
Ultra-relativistic Quantum Molecular Dynamics (UrQMD) transport
model and THERMal heavy-IoN generATOR (THERMINATOR) are applied to
calculate the corresponding $\beta_{t}$ dependence of $\bar{p}/p$
ratio for three gedanken focused isentropic trajectories with
different focusing degree on QCD phase diagram. Finally, we obtained
an observable anomaly in $\beta_{t}$ dependence of $\bar{p}/p$
ratio, which can be employed as a signature of QCD critical point.
\end{abstract}


\maketitle

\section{Introduction}
In recent years, ultra-relativistic high energy heavy ion collisions
(HICs) experiments, such as SPS/CERN
($\sqrt{{s}_\mathrm{NN}}\sim10A$ GeV) and RHIC/BNL
($\sqrt{{s}_\mathrm{NN}}\sim200A$ GeV), have been aimed to search
for the new form of matter, which is composed of deconfinement free
quarks and gluons, and thus called Quark Gluon Plasma (QGP)
\cite{QGP1,QGP2,QGP3,QGP4}. Lattice-QCD calculation predicts that
the phase transition between hadronic and QGP phases at vanishing
baryon chemical potential $\mu_{B}$ is crossover transition whereas
at higher $\mu_{B}$ the transition may become first-order
\cite{lattic1,lattic2}. However, there are many uncertainties for
Lattice-QCD or model calculation to determine the first-order phase
transition boundary as well as the location of corresponding end
point, the so-called QCD Critical Point (QCP) \cite{lattic3}. Thus,
the location and even the existence of the QCP are still open
questions. The uncertainty of theoretical calculation drives us to
explore the properties of hot QCD matter at higher net baryon
density and find experimental evidence of the QCP. Recently, several
experimental programs have planed to search for the QCP, such as the
energy scan program of RHIC/BNL ($\sqrt{{s}_\mathrm{NN}}=5\sim39A$
GeV) \cite{criRHIC1,criRHIC2,criRHIC3}, the light ion program of
NA61 experiment at SPS/CERN ($\sqrt{{s}_\mathrm{NN}}=5\sim17.3A$
GeV) \cite{NA61_1,NA61_2} and also the CBM experiment at FAIR/GSI
($\sqrt{{s}_\mathrm{NN}}\le8.5A$ GeV) \cite{CBM_1,CBM_2,CBM_3}. Many
experimental observables have been also proposed to be the QCP
signatures, such as dynamical fluctuations in $K/\pi$ ratio
\cite{fluc}, two experimental observables correlation \cite{corr},
high order moment of transverse momentum \cite{moment}, {\it etc.}.

In the hydrodynamical description of relativistic heavy ion
collisions, the expansion of the central fireball should be regarded
as isentropic for the negligible entropy production in the latter
evolution, which can be represented as entropy density $s$ to baryon
density $n_{b}$ ratio ($s/n_{b}$) to be constant \cite{isentropic1}.
A trajectory with $s/n_{b}=constant$ for a given colliding system on
the QCD phase diagram is called isentropic trajectory. Recently, the
critical singular properties of QCP have been implemented in
equation of state (EOS) for hydrodynamical description of HICs
\cite{isentropic1,isentropic2} by Asakawa, {\it et al.}. They
pointed out that when the isentropic trajectory passes through the
vicinity of the QCP, it may be deformed-the so-called "QCP focusing
effect" \cite{isentropic1}. They also argued that the QCP focusing
effect may result in an observable anomaly in the $\beta_{t}$
dependence of $\bar{p}/p$ ratio \cite{QCP}, which can be employed as
a robust signature of the QCP. However, whether the focusing effect
can effectively result in an observable anomaly in the $\beta_{t}$
dependence of $\bar{p}/p$ ratio has not been worked out yet; neither
has the corresponding mechanism. In this letter, we will formulate
the QCP focusing effect on the $\beta_{t}$ dependence of $\bar{p}/p$
ratio and apply the UrQMD and THERMINATOR model to calculate the
dependence patterns for three gedanken focused isentropic
trajectories with different focusing degrees on the QCD phase
diagram.

The UrQMD model \cite{UrQMD1} used here is based on the quark,
di-quark, string and hadronic degrees of freedom and relativistic
Boltzmann transport dynamics. It includes 50 different baryon
species(nucleon, hyperon and their resonances up to 2.11 GeV) and 25
different meson species. It is usually used to describe the
freeze-out and breakup of the fireball produced in relativistic
heavy-ion collisions into hadrons. The model has successfully been
applied to reproduce the experimental results from SIS/GSI to
SPS/CERN energies \cite{UrQMD2}. The THERMINATOR model
\cite{Thermal} is a Monte-Carlo event generator designed for
studying of particle production in relativistic heavy ion collisions
from SPS to LHC energies. It implements thermal models of particle
production with single freeze out. The input parameters are those
thermodynamical parameters at freeze out, such as temperature $T$,
baryon chemical potential $\mu_{B}$, {\it etc.}.
\label{Introduction}

\section{Formulation of the QCD critical point focusing effect}

\begin{figure}
\centering
\includegraphics[height=16pc,width=22pc]{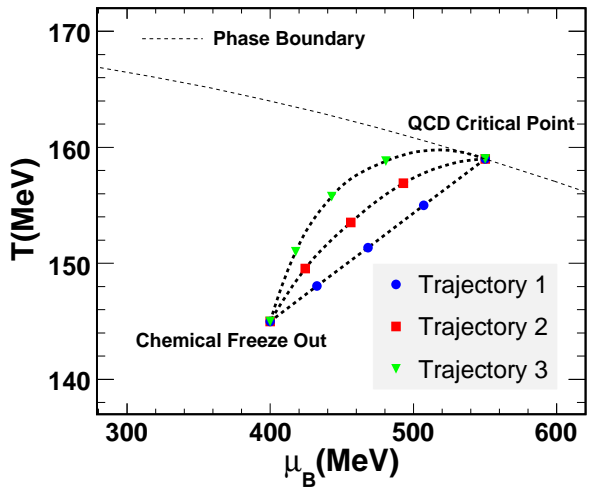}
\caption{Three gedanken isentropic trajectories on the QCD phase
diagram with different focusing degrees. All trajectories meet at
the same QCD critical point (550,159) MeV and chemical freeze out
point (400,145) MeV on the phase diagram.} \label{phase_diagram}
\end{figure}

To quantitatively describe the QCP focusing effect, three gedanken
isentropic trajectories with different focusing degrees, passing
through the common QCP and chemical freeze out point are constructed
on the ($\mu_{B},T$) plane and respectively labeled as Trajectory 1,
2 and 3 in Fig.1. The location of the QCP
($\mu_{c},T_{c}$)=(550,149) MeV chosen here is the same as the Ref.
\cite{QCP}, and the chemical freeze out point
($\mu_{ch},T_{ch}$)=(440,145) MeV is the statistical model fit
result to the data of Pb+Pb 40A GeV fixed target reactions at
SPS/CERN \cite{statistical}. Besides the start and end points, three
other ($\mu_{B},T$) points are also sampled for every trajectory. As
the actual evolution time scale of the isentropic trajectory on the
QCD phase diagram is unknown, the normalized time is used and
defined by the normalized path length, $t=L/L_{tot}$. For each
isentropic trajectory, the $L$ represents the path length along the
trajectory from the considered point to the QCP and $L_{tot}$ is the
length along the trajectory from the chemical freeze out point to
QCP. The colliding system is evolving from the QCP along the
isentropic trajectory to the chemical freeze out point and assumed
to be thermodynamical equilibrium. Then, the $\mu_{B}$ and $T$ of
every sampling point are used as the input thermodynamical
parameters for THERMINATOR model to calculate the corresponding
antiproton and proton numbers along the corresponding isentropic
trajectory on QCD phase diagram.

The normalized time dependence of the antiproton and proton numbers
at the sampled points for the three gedanken isentropic trajectories
are respectively shown in Fig.2, in which the dash lines are the
corresponding 3rd-order polynomial fitting results. It is found in
Fig.2 that QCP focusing effect results in a sharp decrease of proton
number and a non-monotonous increase of antiproton number along the
focused isentropic trajectory. We can also see large discrepancy
among the antiproton numbers of the three trajectories, whereas
there is little discrepancy for the proton number. This means the
antiproton is much more sensitive to the focusing degree than
proton. In Fig. 3, the time evolution of $\bar{p}/p$ ratio along the
three gedanken isentropic trajectories and the corresponding
3rd-order polynomial fitting dash lines are shown. The antiproton to
proton number ratio ($\bar{p}/p$) increases monotonously from QCP
($t=0$) to chemical freeze out point ($t=1$), which can be simply
explained as the decreasing value of $\mu_{B}/T$ along each focused
isentropic trajectory \cite{QCP}.

\begin{figure}
\centering
\includegraphics[height=20pc,width=25pc]{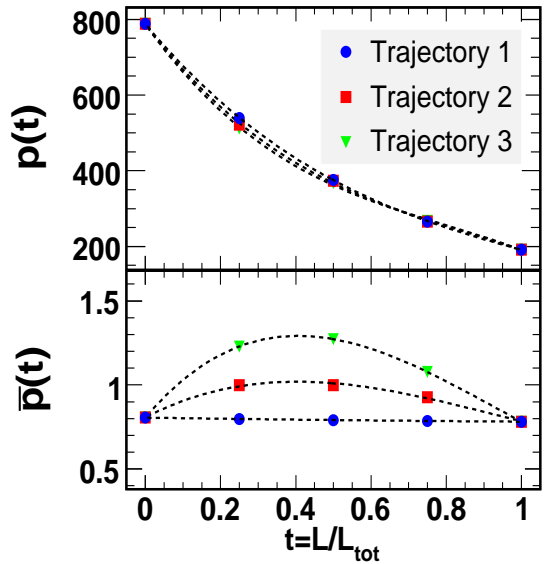}
\caption{The normalized time evolution of the proton(upper panel)
and antiproton (lower panel) numbers along the three gedanken
isentropic trajectories. } \label{pbar_p2}
\end{figure}

Although we have obtained three gedanken focused isentropic
trajectories, the quantitative description of the QCP focusing
effect on $\beta_{t}$ dependence of $\bar{p}/p$ ratio and also the
corresponding mechanism are still ambiguous. In qualitative
analysis, particles with large $\beta_{t}$ would be emitted earlier
from the fireball created in HICs, since the mean free path
generally grows with increasing hadron momentum and becomes
comparable with the fireball size \cite{QCP}. In Fig. 4, the
transverse velocity ( $\beta_{t}$ ) dependences of the average
emission time ( $<t_{emission}>$ ) for antiproton and proton in
Pb+Pb 40 AGeV fixed target reactions are calculated by UrQMD. It is
implied that the antiproton and proton, which are dynamically
emitted from the fireball, would show strong $\beta_{t}-t$
anti-correlation during the cooling down of the colliding system.
The general $\beta_{t}-t$ anti-correlation pattern cannot be
reproduced by hydrodynamic inspired model for their particular
particle freeze out mechanism. The significant increase of
$\bar{p}/p$ ratio with the normalized time $t$ shown in Fig. 3,
together with the strong $\beta_{t}$-$t$ anti-correlation indicates
the experimental observable, $\beta_{t}$ dependence of $\bar{p}/p$
ratio, should be enhanced in the lower $\beta_{t}$ region and
suppressed in the higher $\beta_{t}$ region \cite{QCP}.

\begin{figure}
\centering
\includegraphics[height=15pc,width=21pc]{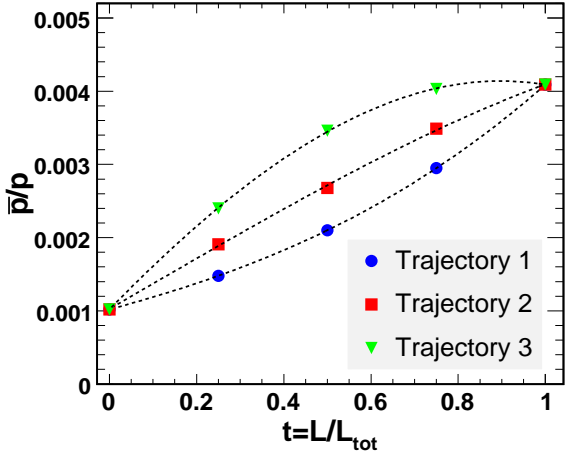}
\caption{The normalized time evolution of the antiproton to proton
ratio for three gedanken isentropic trajectories.} \label{pbar_p}
\end{figure}

The formulation of the QCP focusing effect on $\beta_{t}$ dependence
of $\bar{p}/p$ ratio is based on two simple assumptions: one is that
the numbers of antiproton and proton evolve independently along the
isentropic trajectories and the other is that the numbers of
antiproton and proton emitted at time $t$ are proportional to their
corresponding total number ($\bar{p}(t)$, ${p}(t)$). The two
assumptions can be formulated as:
$$ \bar{p}(t)=k_{1}\times\int{N_{\bar{p}}(\beta_{t},t)}\mathrm{d}\beta_{t} \eqno{(1)}$$
$$ {p}(t)=k_{2}\times\int{N_{{p}}(\beta_{t},t)}\mathrm{d}\beta_{t} \eqno{(2)}$$
,where the $N_{\bar{p}}(\beta_{t},t)$ and $N_{{p}}(\beta_{t},t)$ are
the two-dimensional $\beta_{t}-t$ distributions of the antiproton
and proton, respectively. The two $\beta_{t}-t$ distributions
indicate antiproton and proton are emitted from the colliding system
with a finite probability for certain emission time $t$ and
transverse velocity $\beta_{t}$. The $k_{1}$ and $k_{2}$ are the two
unknown constant coefficients. In addition to equ.(1) and (2), we
also have the boundary condition:
$$ \bar{p}(1)=k_{1}\times\int{N_{\bar{p}}(\beta_{t},1)}\mathrm{d}\beta_{t}=\int^1_0\int{N_{\bar{p}}(\beta_{t},t)}\mathrm{d}\beta_{t}\mathrm{d}t \eqno{(3)}$$
$$ {p}(1)=k_{2}\times\int{N_{{p}}(\beta_{t},1)}\mathrm{d}\beta_{t}=\int^1_0\int{N_{{p}}(\beta_{t},t)}\mathrm{d}\beta_{t}\mathrm{d}t \eqno{(4)}$$
, which means the numbers of antiproton and proton at chemical
freeze out point ($t=1$) are equal to their corresponding sum of
emitted numbers along the isentropic trajectory.

The two constants $k_{1}$ and $k_{2}$ can be determined by
performing integral over the normalized time $t$ on both sides of
equ. (1) and (2), respectively as:
$$k_{1}=\frac{\int^1_0  \bar{p}(t)\mathrm{d}t}{\int^1_0 \int N_{\bar{p}}(\beta_{t},t)\mathrm{d}\beta_{t}\mathrm{d}t }
=\frac{\int^1_0  \bar{p}(t)\mathrm{d}t}{\bar{p}(1)} \eqno{(5)} $$
$$k_{2}=\frac{\int^1_0  {p}(t)\mathrm{d}t}{\int^1_0 \int N_{{p}}(\beta_{t},t)\mathrm{d}\beta_{t}\mathrm{d}t }
=\frac{\int^1_0  {p}(t)\mathrm{d}t}{{p}(1)} \eqno{(6)} $$ We
introduce $D_{\bar{p}}(t)$ and $D_{{p}}(t)$ as the numbers of
antiproton and proton emitted at time $t$ along the isentropic
trajectory, respectively. Then, with equ. (1), (2), (5) and (6), we
have:
$$D_{\bar{p}}(t)=\int{N_{\bar{p}}(\beta_{t},t)}\mathrm{d}\beta_{t}=\frac{\bar{p}(t)}{k_{1}}
=\frac{\bar{p}(t)}{\int^1_0  \bar{p}(t)\mathrm{d}t}\times\bar{p}(1)
\eqno{(7)}$$
$$D_{{p}}(t)=\int {N_{{p}}(\beta_{t},t)}\mathrm{d}\beta_{t}=\frac{{p}(t)}{k_{2}}
=\frac{{p}(t)}{\int^1_0  {p}(t)\mathrm{d}t}\times {p}(1)
\eqno{(8)}$$ It is found that the $D_{\bar{p}}(t)$ and $D_{{p}}(t)$
are only determined by the normalized time $t$ dependence of
antiproton and proton numbers along the isentropic trajectory,
respectively, which are calculated by THERMINATOR model in Fig. 2.

\begin{figure}
\centering
\includegraphics[height=15pc,width=21pc]{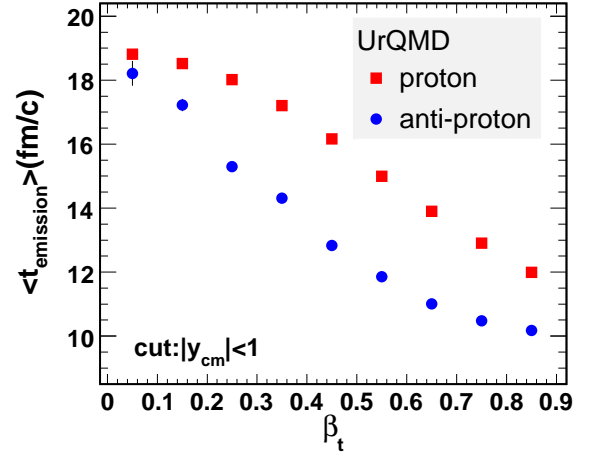}
\caption{The average emission time $<t_{emission}>$ dependence of
the transverse velocity $\beta_{t}$  calculated by UrQMD for Pb+Pb
40 AGeV fixed target reactions at mid-rapidity. } \label{t_cor}
\end{figure}

The experimental observable, $\beta_{t}$ dependence of $\bar{p}/p$
ratio, can be calculated as:
$$\frac{\bar{p}(\beta_{t})}{p(\beta_{t})}=\frac{\int^1_0 N_{\bar{p}}(\beta_{t},t)\mathrm{d}t}{\int^1_0 N_{{p}}(\beta_{t},t)\mathrm{d}t} \eqno{(9)}$$
, which depends strongly on the $\beta_{t}-t$ distributions of
antiproton and proton. Although, the two $\beta_{t}-t$ distributions
are not exactly known, the anti-correlation between $\beta_{t}$ and
$t$ is well known.

To quantitatively analyze and set the benchmark for the $\beta_{t}$
dependence of $\bar{p}/p$ ratio with the QCP focusing effect, we
extract the $\beta_{t}-t$ distributions of antiproton
($N^{U}_{\bar{p}}(\beta_{t},t)$) and proton
($N^{U}_{{p}}(\beta_{t},t)$) from the Pb+Pb 40 AGeV fixed target
reactions implemented by UrQMD. The emission time $t$ here has been
normalized. With the two $\beta_{t}-t$ distributions, the
corresponding $\beta_{t}$ dependence of $\bar{p}/p$ ratio is
calculated by equ.(9) and shown in Fig.5. As QGP phase transition
and QCP have not been implemented in UrQMD model, the monotonous
increase pattern with $\beta_{t}$ up to 0.7 in Fig. 5 is thought to
be normal and without suffering from QCP focusing effect.

\begin{figure}
\centering
\includegraphics[height=15pc,width=21pc]{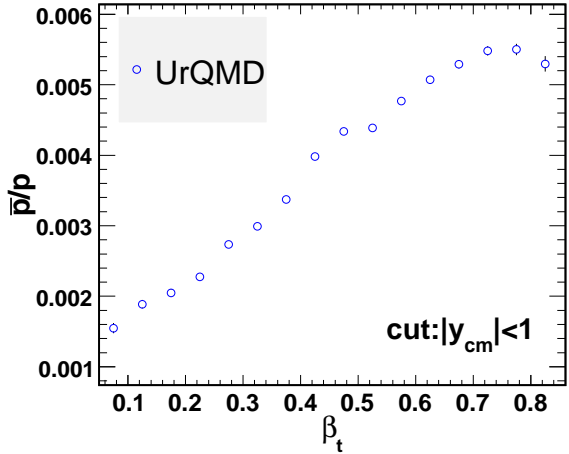}
\caption{ $\beta_{t}$ dependence of $\bar{p}/p$ ratio calculated by
UrQMD for Pb+Pb 40 AGeV fixed target reactions at mid-rapidity.}
\label{ratio_1}
\end{figure}

In order to obtain the anomalous $\beta_{t}$ dependence of
$\bar{p}/p$ ratio resulted from the QCP focusing effects, we modify
the $\beta_{t}-t$ distributions of antiproton
($N^{U}_{\bar{p}}(\beta_{t},t)$) and proton (
$N^{U}_{{p}}(\beta_{t},t)$), which has been calculated by UrQMD
model. The $t$ distributions in the $N^{U}_{\bar{p}}(\beta_{t},t)$
and $N^{U}_{{p}}(\beta_{t},t)$ are replaced by the distributions
$D_{\bar{p}}(t)$ and $D_{{p}}(t)$ derived from equ.(7) and (8),
respectively. For the three gedanken focused isentropic trajectories
in Fig. 1, the resulted $\beta_{t}-t$ distributions for antiproton
and proton, which have been introduced in QCP focusing effect, can
be calculated as:

$$ N^{QCP}_{\bar{p}}(\beta_{t},t)=\frac{N^{U}_{\bar{p}}(\beta_{t},t) }{\int N^{U}_{\bar{p}}(\beta_{t},t) \mathrm{d}\beta_{t}} \times D_{\bar{p}}(t) \eqno{(10)}$$

$$ N^{QCP}_{{p}}(\beta_{t},t)=\frac{N^{U}_{{p}}(\beta_{t},t) }{\int N^{U}_{{p}}(\beta_{t},t) \mathrm{d}\beta_{t}} \times D_{{p}}(t) \eqno{(11)}$$

Consequently, the QCP focusing effect has been implemented in the
$\beta_{t}-t$ distributions for both antiproton and proton through
the equ.(10) and (11), respectively. Thus, the $\beta_{t}$
dependence of $\bar{p}/p$ ratio with QCP focusing effect can be also
calculated by equ.(9) with the modified $\beta_{t}-t$ distributions
( $ N^{QCP}_{\bar{p}}(\beta_{t},t)$, $N^{QCP}_{{p}}(\beta_{t},t)$ ).
The results for the three focused gedanken isentropic trajectories
on the QCD phase diagram are illustrated in Fig. 6. Comparing the
results of UrQMD calculation in Fig. 5 and the results with QCP
focusing effect in Fig. 6, we find that they demonstrate completely
opposite dependence patterns, which indicates the QCP focusing
effect could actually result in an observable anomaly in the
$\beta_{t}$ dependence of $\bar{p}/p$ ratio. In Fig.6, it is also
noticed that the higher focused degree of isentropic trajectory
leads to the steeper $\beta_{t}$ dependence of $\bar{p}/p$ ratio.
Obviously, the anomalous $\beta_{t}$ dependence of $\bar{p}/p$ ratio
can be employed as a sensitive QCP signature.

\begin{figure}
\centering
\includegraphics[height=15pc,width=21pc]{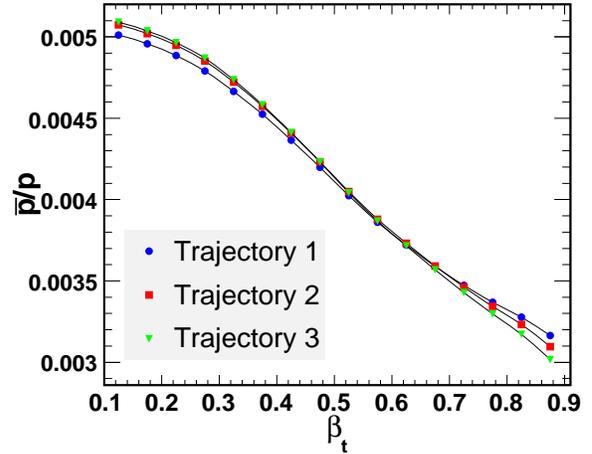}
\caption{$\beta_{t}$ dependence of $\bar{p}/p$ ratio for three
gedanken isentropic trajectories. The solid lines are used to guide
eyes.} \label{ratio}
\end{figure}

\section{Summary and Discussion}
We have formulated the QCP focusing effect on $\beta_{t}$ dependence
of $\bar{p}/p$ ratio with some reasonable assumptions. The
quantitative calculations for the three gedanken focused isentropic
trajectories have been made with UrQMD and THERMINATOR models. In
fact, the real $\beta_{t}-t$ correlation pattern for antiproton and
proton of a colliding system in HICs cannot be measured
experimentally, which directly determines the $\beta_{t}$ dependence
of $\bar{p}/p$ ratio. Therefore, the UrQMD model is used to provide
$\beta_{t}-t$ correlation pattern and the THERMINATOR model is
applied to calculate the normalized time dependence of the
antiproton and proton numbers along the isentropic trajectories.
Then, the anomalous $\beta_{t}$ dependence of $\bar{p}/p$ ratios
have been obtained for three gedanken focused isentropic
trajectories, which means the QCP focusing effect can efficiently
result in an observable anomaly in the $\beta_{t}$ dependence of
$\bar{p}/p$ ratio. We argue that when the isentropic trajectory of
the colliding system on the QCD phase diagram is passing through the
vicinity of the QCP and deformed by QCP focusing effect, an
observable anomaly in $\beta_{t}$ dependence of $\bar{p}/p$ ratio,
enhanced in low $\beta_{t}$ and suppressed in high $\beta_{t}$
region, will be observed. This anomaly may also be reflected in the
$p_{T}$ spectra of antiproton and/or proton. The existing antiproton
$p_{T}$ spectrum for Pb+Pb 40 AGeV fixed target collisions measured
by NA49 collaboration at SPS/CERN exhibits a steeper exponential
slope \cite{QCP,spectrum}. Finally, we propose that it may be
helpful to extract the anomalous structures from the $p_{T}$
spectrum of antiproton and/or proton by performing inverse Laplace
transform on the spectrum \cite{laplace}. The excitation function of
structure variable, which should be predefined, can be useful to
search for the QCP.
\section{Acknowledgement}

This work is supported by the National Natural Science Foundation of
China (10835005, 10675111, 10775131).

\end{document}